\documentclass[pre,floatfix]{revtex4}
\usepackage{graphicx}
\usepackage{graphics}
\usepackage{latexsym}
\usepackage{amsmath, amsthm, amssymb, wasysym}
\usepackage{xcolor}
\usepackage{epsfig}
\usepackage{soul}
\usepackage{cancel}
\usepackage{hyperref}

\begin{document}

\newcommand{\fr}{\frac}
\newcommand{\tl}{\tilde}
\newcommand{\lr}{\langle}
\newcommand{\rl}{\rangle}
\newcommand{\dd}{\partial}

\title{Restarts delay escape over a potential barrier}

\author{R. K. Singh}
\email{rksinghmp@gmail.com}
\affiliation{Department of Biomedical Engineering, Ben-Gurion University
of the Negev, Be'er Sheva 85105, Israel}

\begin{abstract}
In the barrier escape problem, a random searcher starting at the energy
minima tries to escape the barrier under the effect of thermal fluctuations.
If the random searcher
is subject to successive restarts at the bottom of the well, then its escape
over the barrier top is delayed compared to the time it would take in absence
of restarts. When restarting at an intermediate location, the time required
by the random searcher to go from the bottom of the well to the restart
location should be considered. Taking into account this time overhead, we
find that restarts delay escape, independent of the specific nature
of the distribution of restart times, or the location of restart, or the specific
details of the random searcher. For the special
case of Poisson restarts, we study the escape problem for a Brownian particle
with a position-dependent
restart rate $r(x)\theta(x^p_0-x)$, with $x^p_0$ being the location of restart.
We find that position-dependent restarts delay the escape as compared to Kramers
escape time.
We also study ways of modifying time overheads which help expedite escape under
restarts.
\end{abstract}

\maketitle

\section{Introduction}
Thermally activated barrier escape has been a paradigm of chemical
reactions since first formalized by Kramers \cite{kramers1940brownian}.
The rate theory proposed by Kramers showed that the thermally activated
rate of escape decreases exponentially with increasing barrier height
\cite{arrhenius1889dissociationswarme,hanggi1990reaction,
mel1991kramers}. Interestingly, once the temperature of the heat bath
is fixed, the reaction rate is known apriori. This imposes a constraint
on the rate of escape, whereas for almost all practical purposes,
controlling this rate is an important requirement.

Barrier fluctuations provide an important method to control the rate of
escape as it exhibits a local maxima at intermediate rates of barrier
fluctuations \cite{doering1992resonant,bier1993matching,reimann1998reaction}.
In a recent work on Brownian escape it was shown that fine-tuned barrier profiles
can result in significantly enhanced escape rates \cite{chupeau2020optimizing}.
A completely different concept which has rapidly emerged over the past decade and
provides an alternative means to control the rate of escape is the idea
of stochastic restarts \cite{evans2020stochastic,gupta2022stochastic}.
While the previous two methods, namely barrier fluctuations
and profile shaping are specific to thermally activated escape, a stochastic
process subject to restarts is a much broader concept \cite{evans2011JPA,
reuveni2016optimal,eliazar2020mean,pal2016diffusion,
bhat2016stochastic,chechkin2018random,evans2018run,gupta2014fluctuating,
majumdar2015dynamical,singh2022general,masoliver2019telegraphic,nagar2016diffusion,
domazetoski2020stochastic,singh2021backbone,ahmad2019first,falcao2017interacting,
bertin2022stochastic,evans2022exactly,pal2019first,gupta2019stochastic,magoni2020ising,
rotbart2015michaelis,singh2022capture,singh2023bernoulli,
christou2015diffusion,durang2019first,mendez2022nonstandard,
belan2018restart,capala2021dichotomous}. One of the primary advantages brought
about by stochastic restarts is that it renders mean search time finite
even in an unbounded domain \cite{evans2011diffusion}, thus, making it
a method of choice if the aim is to expedite the completion of a stochastic
process \cite{pal2017first,pal2022inspection,roy2024queues,cantisan2024energy}.
Does it mean that stochastic restarts provide a definitive
advantage for the barrier escape problem? While an answer to this question
was provided as an extensive numerical study \cite{cantisan2021stochastic},
a general analysis still eludes us. Moreover, a study of the barrier escape problem
under restarts paves way to a more general scenario, as it would allow us
to compare the relative of advantage of stochastic restarts over, say, barrier
fluctuations. In other words, if we are looking
for a means to speed-up a chemical reaction, should we employ
barrier fluctuations or stochastic restarts?
We answer this question by showing that restarts, in general,
delay escape over a potential barrier.

This also implies that even though restarts provide a certain advantage
for search in infinite domains, it may not be as advantageous to
restart a search process when search is restricted to bounded domains.
In order to establish the above stated result, we study a random searcher
diffusing in a potential $U \equiv U(x)$ in an interval $[0,L]$,
such that $U(0) = 0$ and $U(x)$ is monotonically non-decreasing.
For this simple system to mimic a chemical
reaction, we assign a reflecting wall at $x = 0$ while the wall at $x = L$
is absorbing. This implies that $x = 0$ is the location of the energy minima
and once the diffusing particle starting at the energy minima reaches the energy
maxima at the
absorbing wall, we say that the reaction is complete. The rate of this chemical
reaction is the inverse of the mean escape time (MET), that is, the average
time to go from the bottom of potential well to the top, and for a Brownian
particle it reads \cite{gardiner1985handbook,risken1996fokker}:
\begin{align}
\label{met}
\lr T_{0,L} \rl = \fr{1}{D}\int^L_0 dy~e^{U(y)/D}\int^y_0 dw~e^{-U(w)/D}.
\end{align}
Now if the MET for the Brownian particle to go from bottom of the well to the barrier
top when barriers are fluctuating at a fixed rate is $\lr T^f_{0,L} \rl$, then for
a range of rates: 
\begin{align}
\label{mfpt_br}
\lr T^f_{0,L} \rl \leq \lr T_{0,L} \rl  
\end{align}
holds true \cite{doering1992resonant,bier1993matching,reimann1998reaction}.
Can a similar advantage be achieved via restarts such that the MET under restarts is
lower than $\lr T_{0,L} \rl$ ?

\section{Restarts and associated overheads}
Let us first consider the case when the motion of a random diffusing
particle is restarted from
the bottom of the potential well at $x = 0$. Let $R$ denote the time-interval
of restart, and it can either be a random variable or a deterministic
quantity. Define $\lr T^R_{0,L} \rl$ as the MET
of the random searcher under restarts. Then every time an amount of time $R$ passes,
the searcher is put back to the bottom of the well. This
creates a current from every $x \in (0,L)$ towards the origin which competes
with the natural tendency of the particle to move towards the top and escape.
Meanwhile the clock is ticking, and in the simplest case of exactly one restart
event: $T^R_{0,L} = T_{0,L} + R$ (since restart location is $x = 0$).
As the number of restart events is a non-negative random variable,
this relation implies
\begin{align}
\label{inqTR}
\lr T^R_{0,L} \rl \geq \lr T_{0,L} \rl.
\end{align}

The inequality in (\ref{inqTR}) also implies that the location of restart
$x_0$ should be away from the bottom of the potential well, that is,
$x_0 \in (0,L)$. As restarts bring an advantage by removing trajectories
moving away from the target, there would exist a critical location $x_{0,c}$
such that the mean time taken by the diffusing particle to go from $x = x_0$
to the barrier top in presence of restarts would be identical to the MET for
the same interval without restarts, that is,
\begin{align}
\label{x0c}
\lr T^R_{x_{0,c},L} \rl = \lr T_{x_{0,c},L} \rl,
\end{align}
provided $R$ is sufficiently large. The largeness of $R$ ensures that
the escape takes place with only a few restart events.
It follows from Eq.~(\ref{x0c}) that if $x_0 \in (0,x_{0,c})$
then $\lr T^R_{x_0,L} \rl >
\lr T_{x_0,L} \rl$, that is, restarts are detrimental to escape,
while $x_0 \in (x_{0,c},L)$ implies that $\lr T^R_{x_0,L} \rl < \lr
T_{x_0,L} \rl$, meaning restarts are advantageous. It is to be further noted
that even though Eq.~(\ref{x0c}) is useful only when $R$ is such that
escape under restart is driven by only a few restart events,
existence of the critical location $x_{0,c}$ is conditioned
on the fact that $R$ possesses finite moments, that is,
at least $\lr R \rl$ exists \cite{nagar2016diffusion}.
The exact value of $x_{0,c}$, however, depends on the specific details of the
distribution of restart times and the dynamical properties of the random searcher.
For example, for a Brownian particle subject to Poisson
restarts, Eq.~(\ref{x0c}) leads to the condition $\text{CV}(x_{0,c}) = 1$,
where CV is the coefficient of variation of first passage times in
absence of restarts \cite{pal2017first}.

Now coming back to our problem, the random searcher starts at the
bottom of the well at $x = 0$, and following the above discussion,
it is required that the restart location $x_0 > 0$ for restarts to provide
any advantage. As the clock starts ticking from the moment
the motion started from $x = 0$, the time taken by the searcher
to cover the interval $[0,L]$ should take into account the time to go from
$x = 0$ to $x = x_0$, that is, $T_{0,x_0}$. Moreover, the discussion following
Eq.~(\ref{x0c}) implies that restarts can only reduce $\lr T_{x_0,L} \rl$, that
is, the mean time of travel from $x = x_0$ to $x = L$, provided $x_0 > x_{0,c}$.
As a result, a correct estimation of the time of escape for
the searching particle starting at the bottom of the potential well subject
to restarts should take into account the \textit{time overhead} $T_{0,x_0}$
\cite{reuveni2016optimal}. In the context of chemical reactions this would
correspond to the time to bring the reaction coordinate from the minimal
energy state to a state of an intermediate energy. As the reaction proceeds
from the state of minimal energy via an intermediate state to crossing the energy barrier,
the reaction pathway can be decomposed in disjoint sets such
that $[0,L] = [0,x_0] \cup [x_0,L]$. The decomposition naturally arises,
for example, for the Brownian particle
in light of the fact that the integral in Eq.~(\ref{met}) can
be written as a sum: $\int^L_0 = \int^{x_0}_0 + \int^L_{x_0}$, implying
that the MET decomposes as
\begin{align}
\label{decmp}
\lr T_{0,L} \rl = \lr T_{0,x_0} \rl + \lr T_{x_0,L} \rl~\forall~x_0 \in [0,L].
\end{align}
It is to be noted here that while Eq.~(\ref{met}) holds true for a Brownian
searcher, the decomposition in (\ref{decmp}) holds true for the METs
of an arbitrary
random searcher, and follows from the definition of first passage times
\ref{app00}.

Now in presence of restarts, the time taken by the diffusing
particle to escape over the potential barrier subject to restarts
at an intermediate location $x_0$ is:
\begin{align}
\label{decmpr}
T^R_{0,L;x_0} = T_{0,x_0} + I(T_{x_0,L} < R)T_{x_0,L}
+ I(T_{x_0,L}\geq R)(R + T'^R_{0,L;x_0}),
\end{align}
where $I$ is an indicator variable taking value one when its argument is
true and zero otherwise, $R$ denotes the time of restart, and $T^R_{0,L;x_0}$
is the Kramers escape time with restarts (with an independent copy
$T'^R_{0,L;x_0}$) \cite{reuveni2016optimal,pal2017first,
belan2018restart,roy2024queues}. Taking expectation of Eq.~(\ref{decmpr}) we have
\begin{align}
\label{mean}
\lr T^R_{0,L;x_0} \rl &= \fr{\lr T_{0,x_0} \rl}{\lr I(T_{x_0,L} < R) \rl} +
\fr{\lr \text{min}\{T_{x_0,L},R\} \rl}{\lr I(T_{x_0,L} < R) \rl},\nonumber\\
&= \fr{\lr T_{0,x_0} \rl}{\lr I(T_{x_0,L} < R) \rl} + \lr T^R_{x_0,L} \rl,
\end{align}
wherein the first term accounts for the effect of time overheads
under restarts, and $\text{min}\{a,b\}$ provides the smaller amongst $a$ and
$b$. In absence of any time overheads, only the second term
$\lr T^R_{x_0,L} \rl$ survives and has received extensive
study over the past decade \cite{evans2020stochastic}. However,
in the barrier escape problem, the random searcher starts at the
bottom of the well, and hence any assessment of the dis/advantage
of restarts cannot be made solely on the basis of $\lr T^R_{x_0,L} \rl$,
and a complete analysis of Eq.~(\ref{mean}) is needed.
Now, in absence of any restarts $T_{x_0,L} < R$ holds trivially,
thus reducing Eq.~(\ref{mean}) to (\ref{decmp}). However, when the random
particle is subject to restarts, then from Eq.~(\ref{x0c})
it follows that $\lr T^R_{x_0,L} \rl \geq \lr T_{x_0,L} \rl
~\forall~x_0 \leq x_{0,c}$. As a result, for $x_0 \in (0,x_{0,c}]$ we have
\begin{align}
\label{x0leq}
\lr T^R_{0,L;x_0} \rl \geq \fr{\lr T_{0,x_0} \rl}{\lr I(T_{x_0,L} < R) \rl}
+ \lr T_{x_0,L} \rl
> \lr T_{0,L} \rl,
\end{align}
as $\lr I(T_{x_0,L} < R) \rl < 1$. The other limit when $x_0 > x_{0,c}$
is nontrivial as $\lr T^R_{x_0,L} \rl \leq \lr T_{x_0,L} \rl$ for an appropriately
chosen $R$. In the extreme case, if the diffusing particle
is already at the location of the barrier top, then $\lr T^R_{L,L} \rl,~
\lr T_{L,L} \rl = 0$ holds true identically. Using $x_0 = L$ in (\ref{mean}) leads to
$\lr T^R_{0,L;L} \rl = \fr{\lr T_{0,L} \rl}{\lr I(T_{L,L} < R) \rl}
= \lr T_{0,L} \rl$ as $T_{L,L} = 0$, making the argument of the indicator
variable always true. This implies that
\begin{align}
\label{endp}
\lr T^R_{0,L;x_0} \rl \geq \lr T_{0,L} \rl,~\text{for}~x_0 \in \{x_{0,c},L\},
\end{align}
with the equality holding for $x_0 = L$. In other words, the MET
$\lr T^R_{0,L;x_0} \rl$ is at least equal to the Kramers escape time $\lr T_{0,L} \rl$
when the location of restart $x_0$ equals either $x_{0,c}$ or $L$. Moreover, 
the monotonically decreasing behavior of $\lr T^R_{x_0,L} \rl$ with increasing $x_0$
implies that $\lr T^R_{0,L;x_0} \rl \geq \lr T_{0,L} \rl$ holds true for
every value of $x_0 \in (x_{0,c},L)$ as it holds true at the endpoints of the
interval (see Eq.~(\ref{endp})). Combining this with the inequalities in
(\ref{inqTR}), (\ref{x0leq}), and (\ref{endp}) leads to the result
\begin{align}
\label{x002L}
\lr T^R_{0,L;x_0} \rl \geq \lr T_{0,L} \rl~\forall~ x_0 \in [0,L],
\end{align}
independent of the specific nature of the distribution of restart times
or the specific details about the dynamical properties of the random searcher.

This is the main result of this paper, stating that restarts always delay
escape over a potential barrier, notwithstanding the systemic details.
Moreover, the assumption that the potential energy $U(x)$ is monotonically
non-decreasing was made only to bring the problem in context of the barrier
escape problem. However, the inequality in (\ref{x002L}) holds for arbitrary
potentials, even those exhibiting multisability. Let us now look at some
special cases.

\subsection{Brownian particle under Poisson restarts}
\begin{figure}[t]
\centering
\includegraphics[width=0.6\textwidth]{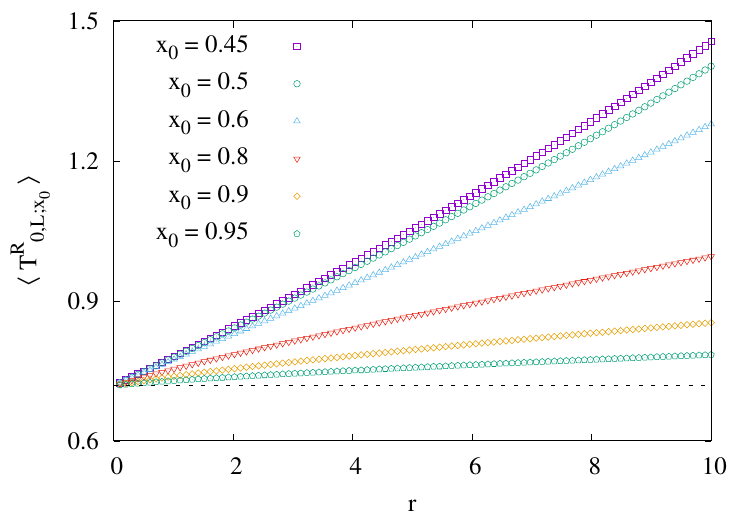}
\caption{Mean time of barrier escape $\lr T^R_{0,L;x_0} \rl$ for
Poisson restarts for different restart locations $x_0$. Black dashed
line represents $\lr T_{0,L} \rl$. Parameter values are $U_0 = 1,~D = 1$,
and $L = 1$.}
\label{fig_mfpt}
\end{figure}
As a representative example, let us consider the escape of a Brownian
particle from a linear potential under Poisson restarts at a rate $r$.
For this special case, Eq.~(\ref{mean}) reduces to \cite{reuveni2016optimal}
\begin{align}
\label{mfptr}
\lr T^R_{0,L;x_0} \rl = \fr{\lr T_{0,x_0} \rl}{\tl{F}(x_0,r)}
+ \fr{1-\tl{F}(x_0,r)}{r\tl{F}(x_0,r)},
\end{align}
where $\tl{F}(x_0,r) = \int^\infty_0 dT~e^{-rT}F(x_0,T)$ is the Laplace
transform of the distribution of first passage times for the Brownian particle starting
at $x = x_0$. For diffusion in the linear potential $U(x) = U_0x$, with
a reflecting wall at $x = 0$ and an absorbing wall at $x = L$, we have
\begin{align}
\label{poiss_fpt}
\tl{F}(x_0,r) = \exp\Big[\fr{U_0}{2D}(x_0-L)\Big]\fr{h_r(x_0)}{h_r(L)},
\end{align}
where
\begin{align}
h_r(w) = \sinh\Big(\fr{U_0w \Delta_r}{2D}\Big) - \Delta_r \cosh\Big(\fr{U_0w \Delta_r}{2D}\Big),
\end{align}
with $\Delta^2_r = 1+\fr{4rD}{U^2_0}$ (\ref{app1}). Furthermore, from
Eq.~(\ref{met}) we have (\ref{app1})
\begin{align}
\lr T_{0,x_0} \rl = \fr{D}{U^2_0}(e^{U_0x_0/D}-1)-\fr{x_0}{U_0}.
\end{align}
Using these values and (\ref{poiss_fpt}) in
(\ref{mfptr}) we find that
\begin{align}
\label{poiss_ineq}
\lr T^R_{0,L;x_0} \rl \geq \lr T_{0,L} \rl~\forall~ r \geq 0~\text{and}~x_0 \in [0,L],
\end{align}
and is described in Fig.~\ref{fig_mfpt}, with the restart locations $x_0$
such that $x_0 > x_{0,c}$ where $x_{0,c} \approx 0.4167$ (\ref{app1}).

\subsection{Poisson restarts with position-dependent rates}
So far we have been considering the scenario in which restart events take place
independent of the location of the random searcher. However, subjecting the
searching particle to restarts when it is close to the top of the barrier takes
it away from the target, bringing in a certain disadvantage. To overcome
this disadvantage, let us consider position-dependent restarts
\cite{pinsky2020diffusive}. Let us further assume that restarts always
bring the random searcher close to the target. This can be achieved,
for example, by defining restart rates as $r(x(t)) \theta(x^p_0-x(t))$, where
$\theta(y) = 1$ if $y \geq 0$ and $0$ otherwise, is the step-function, with
$x^p_0$ being the location of restart. Moreover, this definition ensures that
restarts take place only when the position of the searcher satisfies
$x(t) < x^p_0$.

An application of position-dependent restarts in the present context makes it
evident that the restart location (analogous to the case when restarts are independent
of the position of the searcher) cannot be at the minima of the potential
well at $x = 0$. This is because if $x^p_0 = 0$, then no restart shall take place
in the region $x \in [0,L]$, implying that the MET under position-dependent restart
$\lr T^{R,p}_{0,L;x^p_0} \rl$ equals the Kramers escape time, that is,
\begin{align}
\label{xp00}
\lr T^{R,p}_{0,L;x^p_0} \rl = \lr T_{0,L} \rl,~\text{as}~x^p_0 = 0.
\end{align}
The above result also implies that restarts shall be effective only
when $x^p_0 > 0$. And as discussed above, this would naturally introduce a
time overhead $T_{0,x^p_0}$. The pressing question is, can position-dependent
restarts overcome the disadvantage when restarts take place independent of
the position of the random searcher?

Before we address this question, let us first study the properties of the
critical location $x^p_{0,c}$ associated with position-dependent restarts
and compare it to $x_{0,c}$ defined in
Eq.~(\ref{x0c}), for Poisson restarts. For this purpose, we study the problem in which
a random searcher starts at $x^p_0$, also the restart location, and tries
cross the potential barrier at $x = L$. Now by definition, if restarts depend on the
position such that they always bring the searcher close to the target post restart,
then the MET for position-dependent restarts $\lr T^{R,p}_{x^p_0,L} \rl$ is bounded
above by the MET for the case when restarts take place independent of the position of
the searcher, provided the restart location for both the protocols is $x^p_0$.
This is because in the former case, restarts always bring the searcher close to the
target, while in the latter, restarts can also place the searcher away from the target
when it is in close proximity of the target. As a result,
\begin{align}
\label{pos_MET}
\lr T^{R,p}_{x^p_0,L} \rl \leq \lr T^R_{x^p_0,L} \rl~\forall~x^p_0 > 0.
\end{align}
This implies that position-dependent restarts tend to expand the domain of applicability
of restarts. In other words, if $x_{0,c}$ is the critical location for Poisson restarts,
defined by CV$(x_{0,c}) = 1$ (\ref{app1}), then the critical location for
position-dependent
restarts $x^p_{0,c}$ satisfies the inequality: $x^p_{0,c} \leq x_{0,c}$.
Furthermore, Eq.~(\ref{xp00}) and the inequality in (\ref{pos_MET}) implies that
the critical location for position-dependent restarts is
\begin{align}
x^p_{0,c} = 0.
\end{align}

Now coming back to the question asked in the previous paragraph.
From the above discussion it follows that the critical location
for position-dependent Poisson restarts coincides with the bottom of the well,
and every restart location $x^p_0 > 0$ provides an advantage over Poisson
restarts. Does this mean that position-dependent restarts can provide a
quicker escape compared to escape driven only by thermal fluctuations?
Answer to this question would certainly mean that the associated time
overheads need to be taken into consideration.
To address this, we employ Eq.~(\ref{decmpr}), which for
position-dependent restarts reads:
\begin{align}
\label{decmpr_xp}
T^{R,p}_{0,L;x^p_0} = T_{0,x^p_0} + I(T_{x^p_0,L} < R)T_{x^p_0,L}
+ I(T_{x^p_0,L}\geq R)(R + T'^{R,p}_{0,L;x^p_0}),
\end{align}
with the symbols having their usual meaning as in Eq.~(\ref{decmpr}). Taking the
expectation of the above equation leads to
\begin{align}
\label{mean_xp}
\lr T^{R,p}_{0,L;x^p_0} \rl &= \fr{\lr T_{0,x^p_0} \rl}{\lr I(T_{x^p_0,L} < R) \rl} +
\fr{\lr \text{min}\{T_{x^p_0,L},R\} \rl}{\lr I(T_{x^p_0,L} < R) \rl}.
\end{align}
For Poisson restarts with position-dependent rate $r(x(t))\theta(x^p_0-x(t))$, the
probability that no restart takes place upto time $R$ is $\exp\Big(-\int^R_0 ds~r(x(s))
\theta(x^p_0-x(s)\Big)$, and as a result, the distribution of restart times is
\begin{align}
P(R) = -\fr{d}{dR}\exp\Big(-\int^R_0 ds~r(x(s))\theta(x^p_0-x(s)\Big)
= r(x(R))\theta(x^p_0-x(R))\exp\Big(-\int^R_0 ds~r(x(s))\theta(x^p_0-x(s)\Big).
\end{align}
And if the distribution of first passage times is $F(x^p_0,T)$ then \ref{app2}
\begin{align}
\lr I(T_{x^p_0,L} < R) \rl = \int^\infty_0 dT_{x^p_0,L}~F(x^p_0,T_{x^p_0,L})
\int^\infty_{T_{x^p_0,L}} dR~r(x(R))\theta(x^p_0-x(R))
\exp\Big(-\int^R_0 ds~r(x(s))\theta(x^p_0-x(s)\Big).
\end{align}
The average $\lr \text{min}\{T_{x^p_0,L},R\} \rl$ is estimated analogously \ref{app2}.
This provides us with a means of estimating the MET $\lr T^{R,p}_{0,L;x^p_0} \rl$,
once the form of the spatial dependence of the restart rate $r(x(t))$ is known.

Notwithstanding, a few general properties of the MET $\lr T^{R,p}_{0,L;x^p_0} \rl$ can
be studied. For this purpose, let us rewrite Eq.~(\ref{mean_xp}) in a more recognizable
form:
\begin{align}
\label{mean_xp_ovh}
\lr T^{R,p}_{0,L;x^p_0} \rl = \fr{\lr T_{0,x^p_0} \rl}{\lr I(T_{x^p_0,L} < R) \rl} +
\lr T^{R,p}_{x^p_0,L} \rl,
\end{align}
wherein the second term in Eq.~(\ref{mean_xp}) has been identified as the MET
$\lr T^{R,p}_{x^p_0,L} \rl$ under
Poisson restarts with a position-dependent rate for a random
searcher starting at $x^p_0$. In
Eq.~(\ref{xp00}) we have seen that $\lr T^{R,p}_{0,L;x^p_0} \rl = \lr T_{0,L} \rl$
when the restart location coincides with the bottom of the well, that is, $x^p_0 = 0$.
At the other extreme, when the restart location is at the location of the absorbing
wall at $x^p_0 = L$, then $\lr T^{R,p}_{x^p_0,L} \rl = 0$ identically. This implies
that $\lr I(T_{x^p_0,L} < R) \rl = 1$ as $T^{R,p}_{x^p_0,L} = 0$. And once again,
$\lr T^{R,p}_{0,L;x^p_0} \rl = \lr T_{0,L} \rl$. Summarily,
\begin{align}
\label{mean_xp_endp}
\lr T^{R,p}_{0,L;x^p_0} \rl = \lr T_{0,L} \rl,~\text{for}~x^p_0 \in \{0,L\}.
\end{align}
Furthermore, the smoothness of $\lr T^{R,p}_{0,L;x^p_0} \rl$ in the interval
$x \in [0,L]$ implies that there exists at least one point $x = x^p_m$ at which
the MET is extremum (see Rolle's theorem \cite{courant1965introduction}).
Now the physical interpretation of the terms defining the MET is as follows:
the first term in Eq.~(\ref{mean_xp_ovh}) is the average overhead time in presence
of position-dependent restarts, and the second term is the average time to go
from $x = x^p_0$ to the absorbing wall at $x = L$. Moreover, the former, that
is, $\lr T_{0,x^p_0} \rl/\lr I(T_{x^p_0,L} < R) \rl$, is a monotonically
increasing function of $x^p_0$, while the latter $\lr T^{R,p}_{x^p_0,L} \rl$
is a monotonically decreasing function of $x^p_0$. This implies that
the MET $\lr T^{R,p}_{0,L;x^p_0} \rl$ is a sum of two functions, one of which
is monotonically
increasing while the other is monotonically decreasing over the interval $[0,L]$.
And hence, the MET $\lr T^{R,p}_{0,L;x^p_0} \rl$ is extremum at exactly one point
$x^p_m \in (0,L)$. As a result,
\begin{align}
\label{met_lor}
\lr T^{R,p}_{0,L;x^p_0} \rl > \lr T_{0,L} \rl
\lor \lr T^{R,p}_{0,L;x^p_0} \rl < \lr T_{0,L} \rl~\forall~x^p_0~\in~(0,L).
\end{align}
The mutually exclusive and exhaustive relations asserted between
$\lr T^{R,p}_{0,L;x^p_0} \rl$ and $\lr T_{0,L} \rl$ in the above statement also
imply that if they hold at any one point in the interval $x^p_0 \in (0,L)$,
then they hold true for the whole interval.

To address which of the two relations in the statement (\ref{met_lor}) holds true,
we choose $r(x(t)) = r$ and study motion of a Brownian particle under such
restarts. We shall make use of this later to study the properties of a generic
spatial dependence of the restart rate $r(x(t))$. It is to be noted here that search under
the restart rate $r\theta(x^p_0-x(t))$ has been recently studied 
in Ref.~\cite{tal2024smart} under the name of \textit{smart restarts}. As
the statement in (\ref{met_lor}) holds true $\forall~r > 0$, we
study the MET $\lr T^{R,p}_{0,L;x^p_0} \rl$ for a particular value of restart
rate, namely $r = 1$. We estimate the average escape time by numerically solving a set of
Langevin equations. Time overheads are described by the Langevin equation:
\begin{align}
\label{lang1}
x(t+\Delta t) = x(t) - U_0 \Delta t + \eta(t)\sqrt{\Delta t},~x \in [0,x^p_0],
\end{align}
and once the Brownian searcher reaches the chosen restart location $x^p_0$,
the motion stops. The time taken to cover this interval $[0,x^p_0]$ constitutes
the overhead time $T_{0,x^p_0}$.
For the restart phase, the Brownian motion starts at $x^p_0$ at time $t = T_{0,x^p_0}$
(accounting for the time overhead) and is subject to position-dependent restarts with
$x^p_0$ serving as the restart location. Moreover, the dynamics is subject to restarts
only when $x(t) < x^p_0$ with $t \geq T_{0,x^p_0}$, and is described by the Langevin
equation:
\begin{align}
\label{lang2}
x(t+\Delta t) = \begin{cases}
x(t) - U_0 \Delta t + \eta(t)\sqrt{\Delta t},~\text{with probability}~1-r\Delta t,\\
x^p_0,~\text{with probability}~r\Delta t~\text{and}~x(t) < x^p_0.
\end{cases}
\end{align}
In the Langevin equations (\ref{lang1}) and (\ref{lang2}), $\eta(t)$ is a Gaussian
white noise with mean zero and correlations: $\lr \eta(t_1) \eta(t_2) \rl = 2D\delta(t_1-t_2)$,
with $D$ being the diffusion coefficient.
\begin{figure}[ht]
\centering
\includegraphics[width=0.6\textwidth]{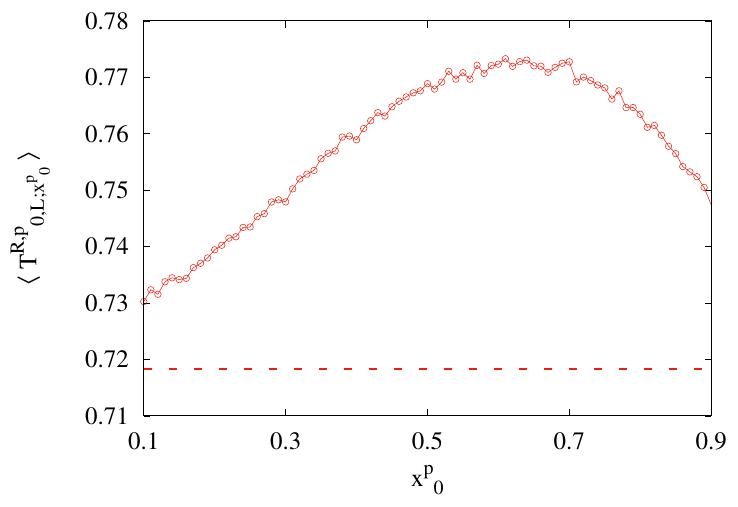}
\caption{MET $\lr T^{R,p}_{0,L;x^p_0} \rl$ for the Kramers escape problem with
the Brownian particle subject to
position-dependent restarts for different restart location $x^p_0$ in presence of
time overheads $T_{0,x^p_0}$. Parameter values $U_0 = 1,~D = 1$, and $L = 1$.
Red dashed line indicates $\lr T_{0,L} \rl$ for these parameter values.
Time-step $\Delta t = 4\times 10^{-5}$ and averaging is performed over
$10^6$ trajectories.}
\label{fig_chem_res}
\end{figure}
Solution of the Langevin Eq.~(\ref{lang1}) for $t \in [0,T_{0,x^p_0}]$
and (\ref{lang2}) for $t \geq T_{0,x^p_0}$ provides us with the MET
$\lr T^{R,p}_{0,L;x^p_0} \rl$ for the Brownian particle starting at the bottom of
the well to cross the barrier top and is described in Fig.~\ref{fig_chem_res}.
The observations in Fig.~\ref{fig_chem_res} implies that the extremum exhibited
by the MET is a maximum. As a result,
\begin{align}
\label{pos_res_ineq}
\lr T^{R,p}_{0,L;x^p_0} \rl > \lr T_{0,L} \rl~\forall~x^p_0~\in (0,L) \land \forall~r > 0.
\end{align}
Apparently, the so called \textit{smart restarts} are no longer smart when applied to
the barrier escape problem and taking into consideration the corresponding time overheads.

For the general case of position-dependent restart rate $r(x(t))\theta(x^p_0-x(t))$,
let us choose the restart location in the neighborhood of the reflecting wall at $x = 0$.
That is, there exists an infinitesimal $\epsilon > 0$ such that $0 < x^p_0 < \epsilon$.
This implies that restarts take place only when the position of the Brownian particle
is in the immediate vicinity of $x = 0$. As a result, the restart rate can be written
as the Taylor expansion:
\begin{align}
r(x) = r(0) + x\fr{d}{dx}r(x)\Big|_{x=0} + \cdots \approx r(0),
\end{align}
as $x$ is close to zero. In other words, in the immediate neighborhood of the reflecting
wall, the restart rate is practically constant. And it follows from the inequality in
(\ref{pos_res_ineq}) that position-dependent restart rates prove detrimental when the
the restart location $x^p_0 \in (0,\epsilon)$. As a result,
the equality in (\ref{mean_xp_endp}) and Rolle's theorem in conjunction with this statement
leads to the result:
\begin{align}
\label{pos_res_ineq00}
\lr T^{R,p}_{0,L;x^p_0} \rl \geq \lr T_{0,L} \rl~\forall~x^p_0~\in [0,L]
\end{align}
and arbitrary spatial dependence of the restart rate $r(x(t))$.

In summary, even though the Poisson restarts with a position-dependent rate $r(x(t))
\theta(x^p_0-x(t))$ provide an advantage over Poisson restarts with a constant rate
$r$, they do not overcome the bounds imposed on the Kramers escape time. This signifies
that restarts cannot be employed to expedite a chemical reaction.

\section{Modification of time overheads}
As is clearly evident from the above discussions, restarts delay escape over a potential
barrier because of the presence of time overheads. Moreover, introduction of restarts
enhances the effect of time overheads (see Eqs.~(\ref{decmpr}) and (\ref{decmpr_xp})).
However, for the purpose of the present analysis, we shall restrict ourselves to the
case of restarts taking place independent of the position of the random searcher.
Now, the primary reason that inclusion of time overheads delays escape when restarts
are introduced is because the time overheads are derived from the same stochastic process
which defines the dynamics of the random searcher. Let us understand this physically.

A random searcher starts at the bottom
of the well, trying to escape by crossing the potential barrier. Once it reaches a
predetermined location $x_0$, the time $T_{0,x_0}$ is measured and the
subsequent motion is now subject to restarts. The MET of the resulting trajectory from
the potential well to the barrier top is given by Eq.~(\ref{mean}).
Furthermore, the inequality in (\ref{x002L}) implies that altering the
restart location $x_0$ would not
be any useful. This leaves us with only one possibility, namely, modifying
the time overhead $T_{0,x_0}$. The pressing question is, if we modify the
time overhead $T_{0,x_0}$ to
$T^h_{0,x_0}$, then can we expect that the inequality in Eq.~(\ref{x002L}) is
reversed? This modification can be achieved, for example, by transferring the
random searcher
from the bottom of the well to the chosen restart location $x_0$ at a
fixed speed. Alternatively, the temperature of the system can be increased resulting
in an increased diffusion coefficient for the motion in the interval $[0,x_0]$.
Both these measures can be chosen so as to reduce the mean overhead time. It is
to be noted here that this modification of the dynamical properties of the searcher is
limited to the interval $[0,x_0]$. Once the random searcher reaches $x = x_0$,
its dynamics proceeds as a random walk subject to restarts with restart location $x_0$.

To understand this concept, let us consider the scenario in which a
Brownian particle is subject to Poisson restarts at a rate $r$.
Furthermore, let us modify the time overheads by increasing the diffusion coefficient
of the Brownian particle. This can be achieved, for example, by
heating the particle while it is diffusing in the sub-interval $[0,x_0)$
(analogous to evaporation from a region \cite{tucci2020controlling}),
and once it reaches $x = x_0$, restoring its diffusion coefficient back to $D$.
Due to increased diffusion coefficient, the Brownian particle would reach $x_0$
in a lesser amount of time, and the modified time overhead $T^h_{0,x_0}$
satisfies $\lr T^h_{0,x_0} \rl \leq \lr T_{0,x_0} \rl$. As a result,
the new MET for the escape process is
\begin{align}
\label{met_mod}
\lr T^h_{0,L;x_0} \rl = \lr T^h_{0,x_0} \rl + \lr T_{x_0,L} \rl,
\end{align}
and is obtained by replacing $\lr T_{0,x_0} \rl$ by $\lr T^h_{0,x_0} \rl$
in Eq.~(\ref{decmp}). However, unlike Eq.~(\ref{decmp}), the result depends
on the value of $x_0$, simply because the dynamics in the two sub-intervals
$[0,x_0]$ and $[x_0,L]$ are different. Moreover, as a consequence of the
modification in the time overheads, the MET under restarts is (see Eq.~(\ref{mfptr}))
\begin{align}
\label{mfptbn}
\lr T^{R,h}_{0,L;x_0} \rl = \fr{\lr T^h_{0,x_0} \rl}{\tl{F}(x_0,r)}
+ \lr T^R_{x_0,L} \rl.
\end{align}
And if restarts are to provide any advantage for the Kramers escape problem
under this modification, then $\lr T^h_{0,x_0} \rl$ should satisfy the
constraint:
\begin{align}
\label{ineq_ovh}
\lr T^{R,h}_{0,L;x_0} \rl \leq \lr T^h_{0,x_0} \rl + \lr T_{x_0,L} \rl,
\end{align}
leading to the upper bound:
\begin{align}
\label{bound}
\lr T^h_{0,x_0} \rl \leq \fr{1}{r}\Big(\fr{\lr T_{x_0,L} \rl}
{\lr T^R_{x_0,L} \rl} - 1\Big).
\end{align}
Positivity of $\lr T^h_{0,x_0} \rl$ implies that
$x_0 > x_{0,c}$ in Eq.~(\ref{bound}). The above inequality also implies that
if $x_0 \leq x_{0,c}$, then no matter what method we choose to modify
the overhead times, restarts will always delay escape over the potential
barrier. However, if $x_0 > x_{0,c}$ then an appropriately modified time
overhead $T^h_{0,x_0}$ can be expected to provide an advantage
with the introduction of restarts.
\begin{figure}[t]
\centering
\includegraphics[width=0.6\textwidth]{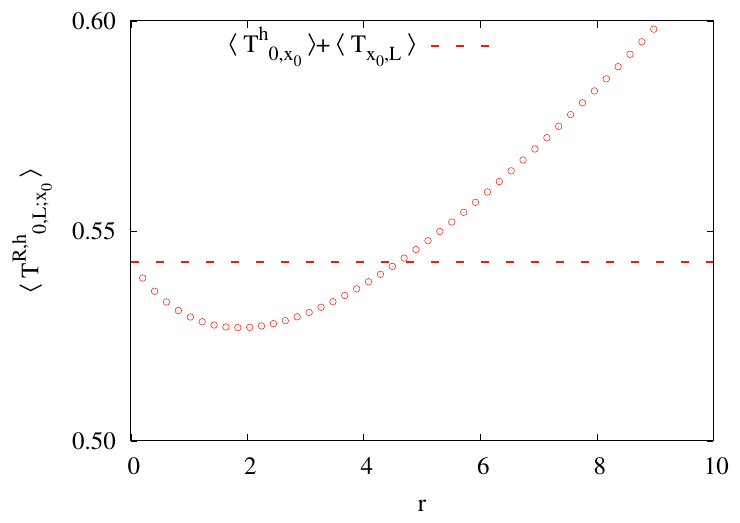}
\caption{Mean time of barrier escape $\lr T^R_{0,L;x_0} \rl$ for
Poisson restarts for restart location $x_0 = 0.7$ and diffusion coefficient
$D_0 = 2$. Parameter values are $U_0 = 1,~D = 1$, and $L = 1$.
}
\label{fig_ovh}
\end{figure}
In the present scenario, we modify the time overheads by effectively heating the
Brownian particle such that it moves in the interval $[0,x_0]$
with an increased diffusion coefficient $D_0$. This provides
a necessary \textit{push} to the particle such that it reaches the
boundary of the domain $x_0 > x_{0,c}$ quickly. Once the Brownian particle
reaches $x = x_0$, the diffusion coefficient is restored to $D$,
and Poisson restarts are introduced for the motion from $x = x_0$ onwards.
With an increased diffusion coefficient $D_0 > D$, the new mean
overhead time is
\begin{align}
\label{ovh_D0}
\lr T^h_{0,x_0} \rl = \fr{D_0}{U^2_0}(e^{U_0x_0/D_0}-1)-\fr{x_0}{U_0},
\end{align}
implying that $\lr T^h_{0,x_0} \rl < \lr T_{0,x_0} \rl$. Using this
in Eq.~(\ref{mfptbn}) we find that increasing
the diffusion coefficient of the Brownian particle can lead to a reduced
MET such that the inequality in (\ref{ineq_ovh}) holds
for a range of restart rates $r$ (see Fig.~\ref{fig_ovh}).
Moreover, the nonmonotonic behavior of $\lr T^{R,h}_{0,L;x_0} \rl$ with varying
$r$ is clearly evident from Fig.~\ref{fig_ovh},
a property characteristic of restarts \cite{reuveni2016optimal,pal2017first}.

The question is, at what cost does the modification of the mean overhead time
$\lr T^h_{0,x_0} \rl$ render application of restarts advantageous?
Answer to this question comes from the fact that when the overheads are
modified, the new MET to the barrier top is given by Eq.~(\ref{met_mod}).
Furthermore, since the diffusion coefficient of the Brownian particle is
$D_0$ for the sub-interval $[0,x_0]$, the Kramers escape takes place with
a new effective diffusion coefficient $D_{eff}$ defined as:
\begin{align}
\label{mfpt_kr}
\lr T^h_{0,L;x_0} \rl = \fr{D_{eff}}{U^2_0}(e^{U_0L/D_{eff}}-1)-\fr{L}{U_0}.
\end{align}
Now using Eqs.~(\ref{ovh_D0}) and (\ref{mfpt_kr}) in (\ref{met_mod})
provides us with the effective diffusion coefficient as:
\begin{align}
\label{deff}
D_{eff}(e^{U_0L/D_{eff}}-1) = U^2_0\Big(\lr T^h_{0,x_0} \rl
+ \lr T_{x_0,L} \rl + \fr{L}{U_0}\Big).
\end{align}
Numerically solving the above equation for $U_0 = 1,~D = 1,~L = 1,~D_0 = 2,~x_0 = 0.7$
(see Fig.~\ref{fig_ovh})
leads to $D_{eff} \approx 1.23$ \cite{press1992numerical}. In other
words, $D_{eff} > D$ implying that for using restarts as a viable strategy
to expedite barrier escape, we need to effectively raise the temperature of the
system. Now, whenever a stochastic process is subject to restarts, an external agency
needs to do work so as to bring back the particle to its restart location
$x_0$ at a fixed rate \cite{fuchs2016stochastic}. In addition, for restarts
to be advantageous, we need to effectively heat the system, which would
incur additional costs. Furthermore, this additional cost is like a necessary
condition, in absence of which restarts always delay escape, following
the inequality in (\ref{x002L}). The question is, if an external agency
needs to do work in order to expedite barrier escape (not to mention the
additional cost), then is it advantageous to employ restarts or assign
fluctuations to the energy barrier?

\section{Discussion}
A general consensus that has emerged from extensive research on stochastic
restarts over the past decade is that the mean time taken by a stochastic process
to reach a threshold can be reduced if the process is subject to restarts.
While this is generically true for search processes in unbounded domains,
searches in bounded domains subject to restarts are expedited only when
specific conditions are met. Taking the practically relevant example of
escape over a potential barrier, we show that employing restarts as a strategy
to expedite barrier escape is bound to fail. This is because when
a random searcher diffusing in a potential landscape is subject to
restarts, time overheads are introduced, and should be taken into consideration
if the MET for the barrier escape is to be correctly estimated.
Moreover, restarting motion over and over again requires external work,
implying that the escape process is better off in absence of this work.

We demonstrate this for the case of a Brownian particle subject to Poisson restarts
with constant and position-dependent rates. By definition, Poisson
restarts with the position-dependent rate $r\theta(x^p_0-x(t))$ are more
efficient compared to Poisson restarts at
the rate $r$, independent of the location of the Brownian searcher.
We show that for a general form of position-dependent restart rate
$r(x(t))\theta(x^p_0-x(t))$, introduction of time overheads enhances the
mean escape time across the potential barrier.
The disadvantage brought about by restarts is overcome by the
introduction of fluctuations in the potential barrier at a fixed rate. This result
is known for over three decades now, namely, barrier fluctuations lead
to a reduction of the MET compared to escape driven only by thermal fluctuations.
This suggests that barrier fluctuations provide an advantage over restarts,
and at this point if an experimenter has to make the choice of restarts
vs barrier fluctuations to expedite the rate of a chemical reaction, we know
for sure the answer.
The question is, are restarts completely useless or some modification of
the restart protocol can lead to a certain advantage?

We answer this question in the affirmative, and find that reducing the overhead
time can result in restarts expediting escape over the barrier.
This reduction of the mean overhead time is, however, achieved by effectively
heating the whole system to an enhanced temperature $D_{eff}$. While we obtain
an explicit relation for this increased temperature for Poisson restarts
in Eq.~(\ref{deff}), the requirement of a necessary \textit{push} like
heating the system is true in general.
Furthermore, even when the mean overhead time can be reduced by
heating the system, the location of restart should be such that $x_0 > x_{0,c}$
as
\begin{align}
\label{boundrr}
\lr T_{0,x_0} \rl \leq \fr{\lr I(T_{x_0,L}<R) \rl}{1-\lr I(T_{x_0,L}<R) \rl}
\lr T^R_{x_0,L}\rl\Big(\fr{\lr T_{x_0,L} \rl}{\lr T^R_{x_0,L} \rl} - 1\Big),
\end{align}
for an arbitrary distribution of restart times and $\lr T_{0,x_0} \rl$ is
positive.

Notwithstanding the speed-up obtained, the additional cost of heating a
section $[0,x_0]$ of the full interval $[0,L]$ raises questions about the
practical constraints on employing restarts. This is especially true in experiments
as it always takes a finite amount time to bring the particle back to its
restart location \cite{tal2020experimental}. For example, how would
the inequality in (\ref{boundrr}) modify for finite time restarts? This
brings us back to the question we asked in the introduction, about applying
restarts to obtain speed-up in the rate of escape. We find that before employing
restarts as a method to expedite escape in any practical scenario, a thorough
assessment of the costs incurred and advantages gained should be made.
To restart, or not to restart, should be the question to ask.

\appendix

\section{Decomposition of first passage times}
\label{app00}
For a random searcher starting at $x_0 \in (0,L)$, and searching for the
target located at the absorbing wall at $x = L$, the first passage time
is defined as:
\begin{align}
\label{def1}
T_{x_0,L} = \text{min}\{T > 0 : x(0) = x_0 \land x(T) = L\}.
\end{align}
Choosing $x_0 = 0$ in the above statement provides us with the escape time
$T_{0,L}$ taken by the random searcher to go from the bottom of the well to the
top of the barrier. Now the motion from $x = 0$ to $x = L$ can be decomposed
in two phases: (i) one in which the random motion starts at $x = 0$ and
ends when the random searcher first reaches $x = x_0$, and (ii) random motion
starts at $x = x_0$ and ends when $x = L$ for the first time. As this
decomposition holds for any $x_0 \in (0,L)$, the definition in (\ref{def1})
leads to
\begin{align}
T_{0,L} = T_{0,x_0} + T_{x_0,L}. 
\end{align}
Taking expectation of both sides leads to Eq.~(\ref{decmp}).

\section{Escape across a potential barrier}
\label{app1}
Consider a Brownian particle with diffusion coefficient $D$ and
diffusing in the linear potential $U(x) = U_0 x$
in the interval $[0,L]$. The backward equation for the survival probability
$q(x,t)$ reads \cite{risken1996fokker,gardiner1985handbook}:
\begin{align}
\label{qxt}
\dd_t q(x,t) = -U'(x)\dd_x q(x,t) + D\dd_{xx}q(x,t),
\end{align}
with a reflecting wall at the bottom of the well and an absorbing wall at the
top. The resulting boundary conditions are: $\dd_x q(x,t)|_{x=0} = 0$ and
$q(L,t) = 0$ with the initial condition $q(x,0) = 1$. Laplace transforming
Eq.~(\ref{qxt}) leads to
\begin{align}
\label{qxs}
\dd_{xx}\tl{q}(x,s) - \fr{U_0}{D} \dd_x \tl{q}(x,s) - \fr{s}{D}\tl{q}(x,s) =
-\fr{1}{D},
\end{align}
where $\tl{q}(x,t) = \int^\infty_0 dt~q(x,t)$ denotes Laplace transform.
The general solution of Eq.~(\ref{qxs}) is
\begin{align}
\tl{q}(x,s) = a_1 e^{m_1 x} + a_2 e^{m_2 x} + \fr{1}{s},
\end{align}
where $m_1 = \fr{U_0}{2D}(1+\Delta_s)$ and $m_2 = \fr{U_0}{2D}(1-\Delta_s)$.
We estimate the constants $a_1$ and $a_2$ using the initial and boundary
conditions resulting in
\begin{align}
\tl{q}(x,s) = -\fr{e^{-U_0L/2D}}{2s}\times\fr{1-\Delta_s}{h_s(L)}
\exp\Big[\fr{U_0 x}{2D}(1+\Delta_s)\Big]
+\fr{e^{-U_0L/2D}}{2s}\times\fr{1+\Delta_s}{h_s(L)}
\exp\Big[\fr{U_0 x}{2D}(1-\Delta_s)\Big]
+\fr{1}{s},
\end{align}
with $h_s(L) = \sinh\Big(\fr{U_0 L}{2D}\Delta_s\Big) - \Delta_s
\cosh\Big(\fr{U_0 L}{2D}\Delta_s\Big)$. As a result, the distribution of
first passage times in Laplace space: $\tl{F}(x,s) = 1-s\tl{q}(x,s)$ reads
\begin{align}
\label{fptd}
\tl{F}(x,s) = \exp\Big[\fr{U_0}{2D}(x-L)\Big]\fr{h_s(x)}{h_s(L)}.
\end{align}
For the case discussed in the paper we have $U_0 = 1,~L = 1,~D = 1$ and as
a result the moments of first passage times are:
\begin{subequations}
\begin{align}
\lr T_{x,1} \rl &= -\lim_{s\to 0}\fr{d}{ds}\tl{F}(x,s) = e-e^x+x-1,\\
\lr T^2_{x,1} \rl &= \lim_{s\to 0}\fr{d^2}{ds^2}\tl{F}(x,s) =
2\Big[-e^{x+1}-2e^x+e^2+2e+(x+1)e^x+(x-3)e
+(x-1)+\fr{(1-x)^2}{2}\Big].
\end{align}
\end{subequations}
As a result CV$(x)=\sqrt{\fr{\lr T^2_{x,1} \rl - \lr T_{x,1} \rl^2}
{\lr T_{x,1} \rl^2}}$ and solving for CV$(x)=1$ leads to $x \approx 0.4167$
\cite{press1992numerical}.

\section{Poisson restarts with a position-dependent rate}
\label{app2}
For position-dependent restart rate $r(x)$ and first passage time
distribution $F(t)$ we have
\begin{align}
\lr I(T<R) \rl &= \int^\infty_0 dR~P(R) \int^\infty_0 dT~F(T) I(T<R)\nonumber,\\
&= \int^\infty_0 dR~r(x(R))\exp\Big(-\int^R_0 ds~r(x(s))\Big)\int^R_0 dT~F(T)\nonumber,\\
&= \int^\infty_0 dT~F(T)\int^\infty_T dR~r(x(R))\exp\Big(-\int^R_0 ds~r(x(s))\Big),
\end{align}
wherein the last step has been obtained by changing the order of integration.
In a similar manner
\begin{align}
\lr I(T<R)T \rl &= \int^\infty_0 dT~TF(T)\int^\infty_T dR~r(x(R))
\exp\Big(-\int^R_0 ds~r(x(s))\Big),\\
\lr I(T\geq R)R \rl &= \int^\infty_0 dT~F(T)\int^T_0 dR~R r(x(R))
\exp\Big(-\int^R_0 ds~r(x(s))\Big).
\end{align}

\bibliography{ref}

\end{document}